\documentclass[twocolumn,english,prl,twocolumn]{revtex4-1}
\usepackage{lmodern}
\usepackage{lmodern}
\usepackage[T1]{fontenc}
\usepackage[latin9]{inputenc}
\usepackage{graphicx}
\usepackage{esint}
\usepackage{upgreek}
\usepackage[dvipsnames]{xcolor}
\makeatletter
\usepackage{verbatim}
\usepackage{amstext}
\usepackage{babel}

\makeatother

\usepackage{babel}

\begin{document}

%\title{Observation of resistance anomaly and and sequential negative differential resistances near superconducting transition in Ag/Au nano-structures}
\title{Observation of excess resistance anomaly at resistive transitions in Ag/Au nanostructures}

\author{Phanibhusan S. Mahapatra$^{1}$ $^{\#}$, Subham Kumar Saha$^{2}$ $^{\#}$, Rekha Mahadevu$^{2}$, Saurav Islam$^{1}$, Pritha Mondal$^{2}$, Shreya Kumbhakar$^{1}$, T. Phanindra Sai$^{1}$, Satish Patil$^{2}$, U. Chandni$^{3}$, Anshu Pandey$^{2}$ \& Arindam Ghosh$^{1,4}$}
\vspace{1cm}
\affiliation{$^{1}$Department of Physics, Indian Institute of Science, Bangalore
560 012, India. }
\affiliation{$^{2}$Solid State and Structural Chemistry Unit, Indian Institute of Science, Bangalore 560012, India }
\affiliation{$^{3}$Department of Instrumentation and Applied Physics,
Indian Institute of Science, Bangalore 560012, India }
\affiliation{$^{4}$Centre for Nano Science $\&$ Engineering, Indian Institute of Science, Bangalore
560 012, India. }

\begin{abstract}
The resistive transition in nanocomposite films of silver (Ag) nanoclusters of $\sim 1$~nm diameter embedded in gold (Au) matrix exhibits an anomalous resistance peak at the onset of the transition, even for transition temperatures as high as 260~K. The maximum value of the resistance ranges between $\sim 30\% - 300\%$ above that of the normal state depending on devices as well as lead configuration within a single device. The excess resistance regime was observed in about $10$~\% of the devices, and extends from $\sim 10- 100$~K. Application of magnetic field of 9~T was found to partially suppress the excess resistance. From the critical current behavior, as well as negative differential resistance in the current-voltage characteristics, we discuss the possibility of interacting phase slip centers and alternate physical scenarios that may cause the excess resistance in our system.
\vskip 0.5cm
\# These authors contributed equally.
\end{abstract}

\maketitle

Occurrence of anomalous resistance peak at the onset of superconducting transition is one of the most ubiquitous phenomenon observed in many disordered superconductors ranging from metals, high $T_C$ ceramics to granular superconductors realized with bottom-up synthesis approach \cite{berlincourt1959hall,grassie1970transition,ems1971resistance,tajima1984magnetic,tajima1984giant,yamamoto1985giant,lindqvist1990new,
crusellas1992giant,santhanam1991resistance,wan1993interlayer,vaglio1993explanation,suzuki1994resistance,moshchalkov1994intrinsic,
mosqueira1994resistivity,park1995resistance,silva1997microwave,park1997resistance,strunk1998resistance,arutyunov1999resistive,buzea2001origin,
wang2007observation,harada2010large,zhang2013metal}. Although a generic explanation remains elusive, appearance of such anomalies is attributed to geometric and/or non-equilibrium effects at the normal-superconductor interfaces in the presence of structural and compositional inhomogeneities \citep{arutyunov1999resistive,park1997resistance,crusellas1992giant,lindqvist1990new}. A transition to vanishingly small electrical resistance state and the concurrent emergence of diamagnetic susceptibility have recently been observed in nanocomposite films of Ag nanoclusters ($\sim 1$~nm diameter) dispersed on Au matrix close to room temperature, suggesting possible emergence of superconductor-like macroscopic coherence at elevated temperatures \cite{thapa2018coexistence}. The current ($I$)-voltage ($V$) characteristics in these films near the resistive transition was  also reported to exhibit critical current behavior with strong hysteresis and discrete resistance steps \cite{islam2019current}. This was attributed to a highly non-equilibrium nature of the quasi-particle transport and filamentary flow of supercurrent through dissipative phase slip centres (PSC) close to the transition. Counterarguments based on weak-links \cite{urazhdin2019comment} and percolative decoupling of current and voltage leads \cite{pekker2018comment} have also been proposed, prompting further experimentation on the nature of current flow in this unique system. In this paper, we report the observation of excess resistance anomaly in Ag/Au nanostructure film in a multi-terminal geometry at the onset of the resistive transition. We also find that below the transition temperature, the $I-V$ characteristics exhibit current-driven sequential voltage-peaks and negative differential resistances (NDR). The anomaly is partially suppressed at high magnetic field, and bears close resemblance to its key characteristics at the conventional normal-superconductor transitions \cite{arutyunov1999resistive,lindqvist1990new,kwong1991interfacial,crusellas1992giant,zhang2013metal}. 

\begin{figure*}[t]
\includegraphics[clip,width=18cm]{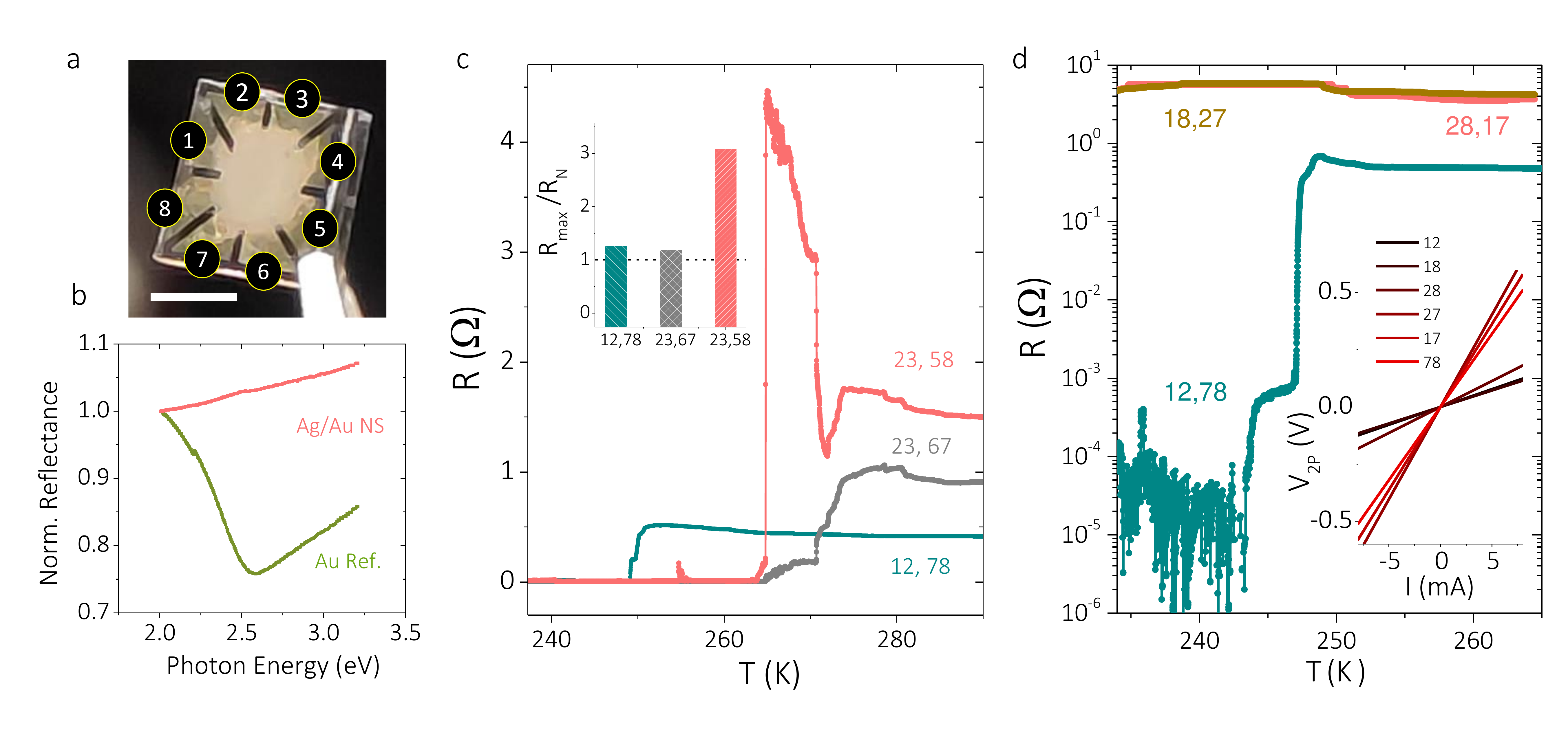}\caption{\textbf{Resistive transition in Ag/Au nanostructure film:} (a) Optical image a typical film with electrical contacts arranged in van der Pauw geometry. The scale bar represents a length of 5~mm. (b) The reflectance spectra (normalized by its value at $2$~eV) is compared for Ag/Au nanostructure film and a reference Au-nanoparticle assembly. (c) The $R-T$ curves for three different regions of the film showing the resistive transition. Above the transition, each current-voltage lead configuration exhibits anomalous increase of $R$. The inset shows the extent of anomalous increase characterized by the ratio of maximum in $R$ and that ($R_N$) at $T \gg T_C$. (d) The $R-T$ plots for different lead configurations between contacts $1,2,7$~and~$8$. The inset shows the pair-wise current-voltage characteristics for various combinations.}
\end{figure*}

The Ag/Au nano-structure films were made by planting Ag nano-particles ($\sim 1$~nm) into a gold matrix using colloidal technique \cite{thapa2018coexistence,islam2019current,thapa2019unconventional}. Briefly, silver clusters were prepared by the reduction of silver nitrate in cetyltrimethylammonium bromide with sodium borohydride. Gold was then introduced in the form of tetrachloroauric acid. Multiple layers of the film ($\approx 100$~nm) were dropcast on pre-patterned $10/100$ nm Cr/Au-electrical contacts on glass substrate arranged in van der Pauw geometry as shown in Fig.~1a. Fig.~1b shows the normalized optical reflectance from a typical flim over the energy range of $2-3$~eV, and compared with that from a reference Au-nanoparticle film. The absence of plasmonic absorption at $\sim 2.6$~eV is one of the key features of the Ag/Au nanostructures. A detailed discussion of this process as well as its optical properties have been reported elsewhere \cite{thapa2018coexistence,thapa2019unconventional}. Fig.~1c shows the temperature dependence of four-probe resistance ($R$) measured with a constant bias current of $1$~mA in different current-voltage contact configurations across the sample. The four-probe $R$ for $23,67$ ($I: 2,3$ and $V: 6,7$) and $23,58$ channels were recorded simultaneously in the same thermal cycle, while that for $12,78$ with $45^\circ$ rotated lead configuration was measured in a separate cool down. The resistance of the as-prepared films varies from $0.4-1.5~\Omega$ across the different regions of the film with corresponding resistivity $\approx 4-15 \times 10^{-8}$~$\Omega$-m \cite{islam2019current}. As evident from Fig.~1c, the resistive transition occurs between $\sim 250 - 265$~K, although both transition temperature ($T_C$) and its nature ({\it e.g.} width) depend strongly on the lead configuration. In some transitions, a residual resistance of $\lesssim 5$~m$\Omega$ at $T < T_C$ was observed, possibly due to the granular nature of the film or weak coupling to the leads, and subtracted manually. We also observed a strong hysteresis between the cooling and the heating cycles (supplementary information), and unless specified otherwise , the data presented here are from the heating cycles. 

In order to verify inter-connectivity between all the leads across the transition, and absence of any unintended constriction due to $T$-dependent structural changes, we have measured ($1$) $R-T$ for various current-voltage lead configurations between $1,2,7$ and $8$, and ($2$) two-probe current-voltage characteristics between all pairwise lead combinations. As shown in Fig.~1d, in spite of nearly $\sim$ four decades of change in resistance in the $R(12,78)$ channel at $\sim 245$~K, the $R(18,27)$ and $R(28,17)$ channels remain essentially unaffected, suggesting (1) absence of any major structural modification in the region between the lead pairs ($1$,$2$) and ($7$,$8$) as a function of $T$, and (2) the physical location of the low resistance/superconducting filament across leads ($1$ and $2$) and/or ($7$ and $8$). The metallic inter-connectivity between the leads is further ensured from linear two-probe $I-V$ characteristics between all lead pairs (inset of Fig.~1d). 

\begin{figure*}[t]
\includegraphics[clip,width=18cm]{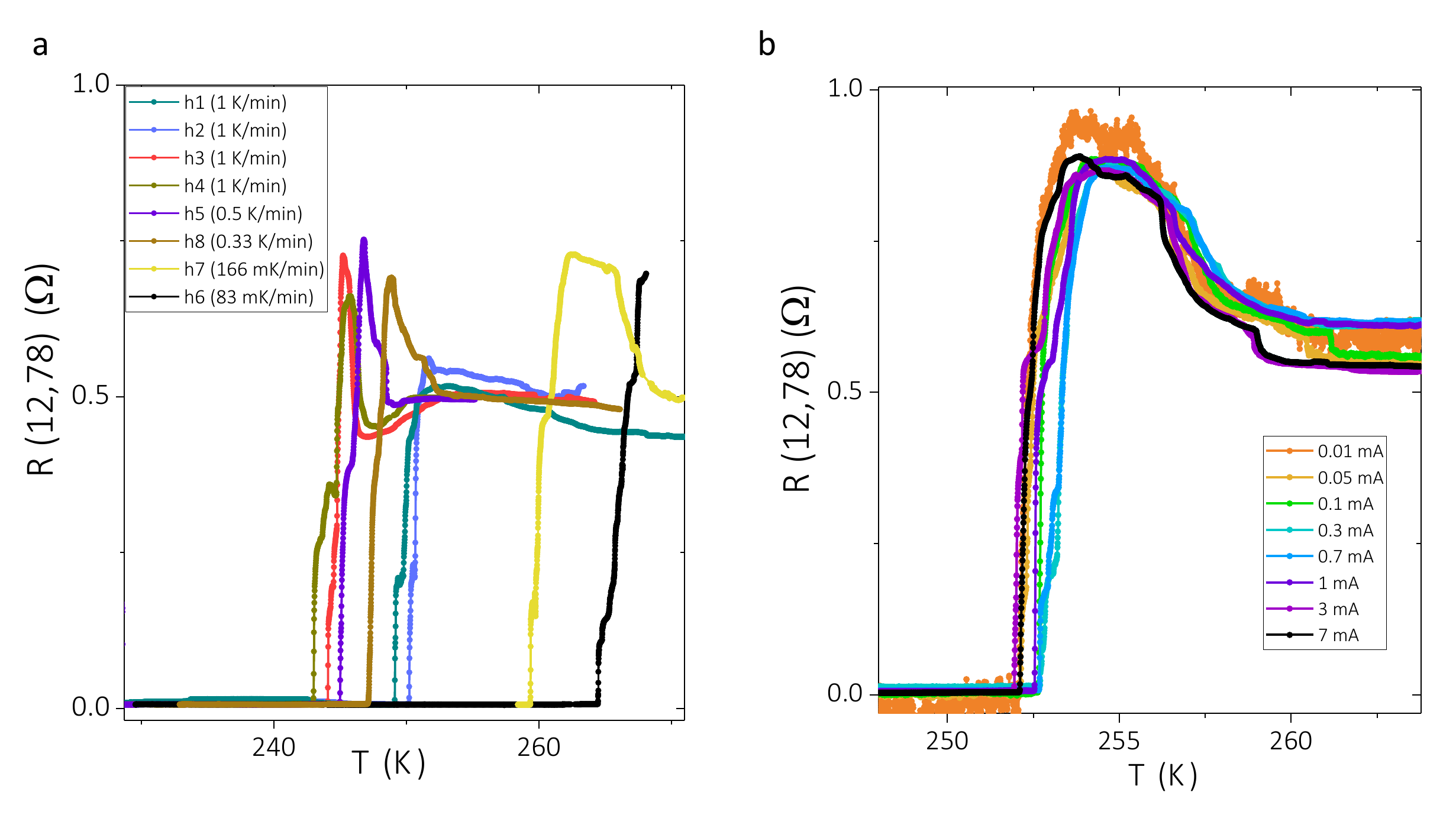}\caption{\textbf{Excess resistance anomaly in resistive transition in Ag/Au nanostructure film:} (a) The temperature-sweep rate dependence of the resistive transition and excess resistance anomaly for eight consecutive thermal cycles with ramp rate varying from $1$~K/min to $83$~mK/min. (b) $R-T$ curves across the transition for eight different DC-current biasing ($10$~$\mu$A - $7$~mA).}
\end{figure*}

In Fig.~1c, $R$ in all three lead configurations exhibits an increase in magnitude $\sim 10 - 20$~K before transition to the low resistance state. Such an increase was observed in $\sim 10$~\% of the devices that showed the resistive transition. This anomalous increase in $R$, characterized by the ratio of maximum in resistance $R_{max}$ close to $T_C$ and that ($R_N$) at $T \gg T_C$ varies between $1.3 - 3$ for different regions of the sample (Fig.~1c inset). The increase was observed irrespective of the AC or DC biasing of current (see supplementary information). The strong variation in the anomaly among different lead configurations is common in known inhomogeneous superconducting systems where it can also be affected by numerous intrinsic and extrinsic parameters, including bias current, thermal history, magnetic field, rf noise and magnetic impurities \citep{arutyunov1999resistive,park1995resistance,strunk1998resistance,crusellas1992giant,lindqvist1990new}.

In order to investigate the thermal history and bias current dependence of excess resistance anomaly (ERA) in our Ag/Au nanostructures, we focus on the $R(12,78)$ channel, which showed reasonably stable and reproducible behavior in multiple thermal cycles. In Fig.~2, we examined the $R-T$ behavior across the transition as a function of both rate of $T$ sweep (Fig.~2a) and biasing current (Fig.~2b). The $R-T$ data in Fig.~2a was taken at a constant bias current of $1$~mA for eight consecutive heating cycles at rate ranging from $83$~mK/min - $1$~K/min. The ERA evolves and eventually stabilizes with progressive thermal cycling with $T_C$ varying by $< 5$~K between the two successive thermal cycles for high sweep rates ($1$~K/min). When the sweep rate is decreased, the ERA remains robust but the $T_C$ increases considerably, indicating the nanostructure film exists in an out-of-equilibrium state. Our earlier experiments \cite{thapa2018coexistence,islam2019current} indicated that $T_C$ in the Ag/Au nanostructures also change with ageing, which may complicate unambiguous determination of the effect of the sweep rate alone. A small 'dip' in $R$ immediately following the peak, observed most clearly for cycles h$3-$h$5$, has also been observed at the superconductor to normal transitions in Au$_{0.7}$In$_{0.3}$ rings \cite{wang2007observation} and high-$T_C$ Bi$_{2}$Sr$_{2}$CaCu$_{2}$O$_{8-y}$ films \cite{wan1993interlayer} that associates the resistance anomaly to the fluctuation regime \cite{lindqvist1990new}. In the eight subsequent thermal cycles (at sweep rate $0.5$~K/min), the biasing current was varied from $10~\mu$A to 7~mA (Fig.~2b). The variation in $T_C = 252.5\pm0.3$~K among these cycles is less than $\sim 0.7$~K, and ERA remains essentially unaffected in spite of nearly three decades of change in current. We note that although few experiments with patterned quasi-1D superconductors report suppression of excess resistance at large biasing current \cite{nordstrom1992resistance,arutyunov1999resistive}, several other investigations, for example those with the high-$T_C$ films \cite{mosqueira1994resistivity}, suggest little or no effect of current density on the anomaly. 

\begin{figure*}[t]
\includegraphics[clip,width=18cm]{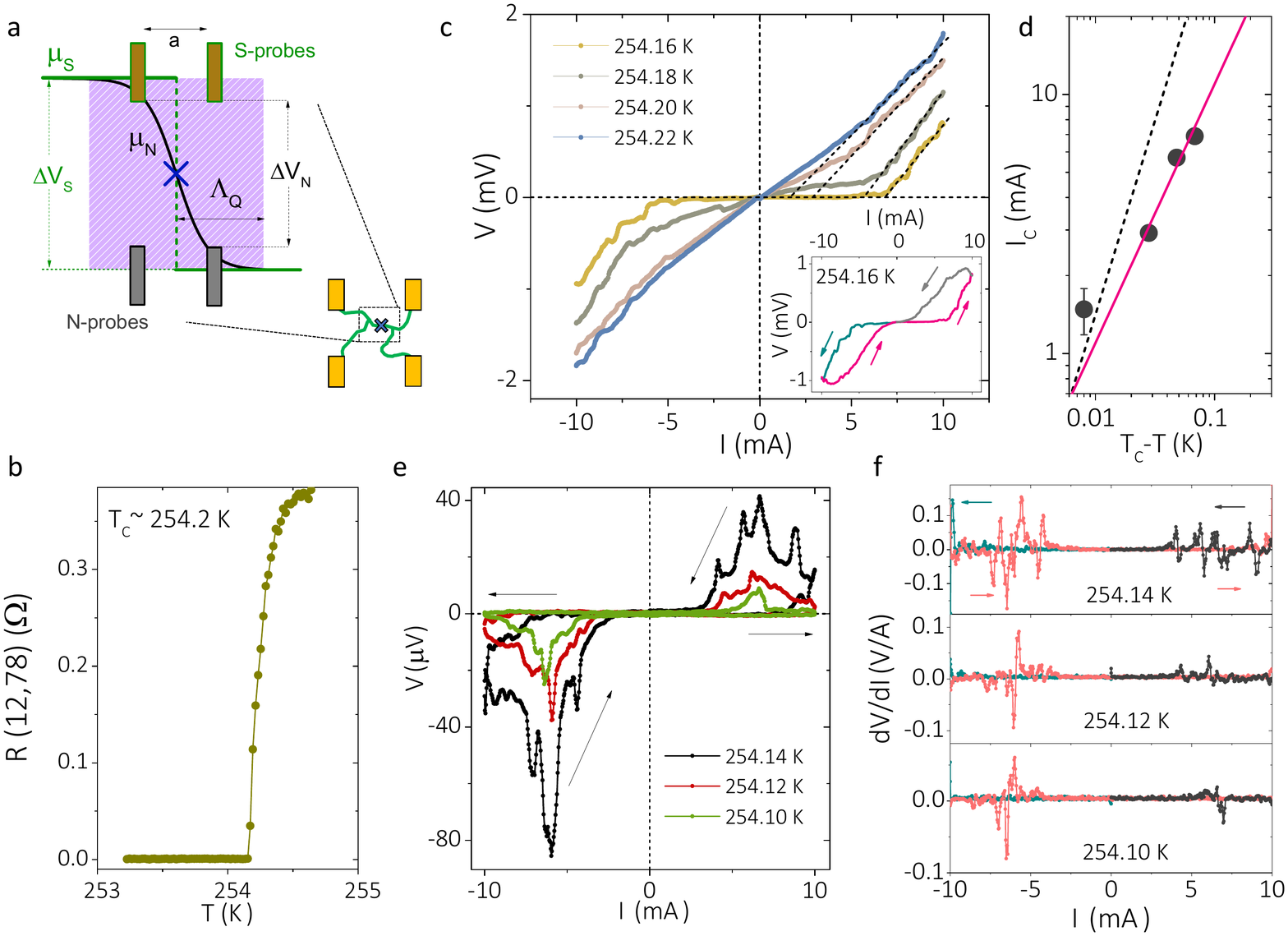}\caption{\textbf{ Current-voltage characteristics}: (a) Schematic showing the filamentary network of the supercurrent carrying paths and the origin of excess resistance anomaly across a phase slip center (PSC).(b) $R-T$ curve obtained from four-probe voltage at $-1$~mA for each $I-V$ cycle.  (c) The four-probe $I-V$ characteristics across the transition for four representative temperatures close to the transition. The critical current ($I_C$) was obtained by the extrapolation of the rapid rise of $V$ (dashed lines) to the current axis. Inset: Full $I-V$ cycle showing hysteresis between the ramping directions. (d) Variation of $I_C$ with temperature. The solid line represents the linear dependence $I_C \propto (T_C-T)^{\beta}$, where $\beta = 1$ and the dash lines corresponds to $\beta = 3/2$. (e) $I-V$ curves below transition for three representative $T$, showing the sharp sequential spikes in $V$. (f) Differential resistance $\frac{dV}{dI}$ obtained from (e) as a function of biasing current.}
\end{figure*}

%As shown in Fig.~2a, the excess resistance ratio, $R/R_N$, the shape and the width of the anomalous peak are found to be dependent of the thermal history i.e, the temperature sweep range, ramp rate etc. From thermal cycle $6-8$, $R$ shows a reduction first as the temperature is lowered before the anomalous peak near $T_C$. But the initial reduction in $R$ gradually disappear in the later thermal cycles. The down-turn in $R$ before the anomaly also suggests that the peak occurs after $T_C$ in the fluctuation regime.

While there is no known analogue of the ERA in non-superconducting metallic nanostructure, the diversity of superconducting devices and materials that exhibit such a phenomenon has led to considerable debate in regard to its microscopic origin. An increase in resistivity in the vicinity of $T_C$ can arise from quasi-particle tunneling between the superconducting-normal (S-N) puddles in case of a granular superconductor \cite{rubin1992observation} or vortex dynamics in the fluctuation regime \cite{francavilla1991observation, harada2010large}. Large excess resistance in granular doped nano-diamond films has been attributed to superconducting transition through an intermediate Bosonic insulating phase \cite{zhang2013metal}. In case of out-of-line probe configuration, such as ours, Vaglio {\it et al.} showed that geometric redistribution of current in the presence spatially varying $T_C$ can provide a simple explanation of the resistance anomaly \cite{vaglio1993explanation}. Non-equilibrium effect at the S-N boundary \cite{arutyunov1999resistive,kwong1991interfacial} and nucleation of interacting phase slip centers (PSC) \cite{park1995resistance,park1997resistance} have also been extensively discussed, particularly in the context of quasi-1D superconducting wires or filaments. In this latter case (see Fig.~3a), the qualitative understanding of the anomaly is based on the argument that when a PSC is formed in the current path and between the (superconducting) voltage leads, the supercurrent ($I_s$) is locally suppressed and the rest of the current is carried as quasi-particle current. The resulting voltage drop across a PSC is given by $V = 2\Lambda_Q \rho_N (I-I_s)/A$, where $\Lambda_Q$, $\rho_N$ and $A$ are the charge imbalance length, normal state resistivity and cross sectional area, respectively \cite{tinkham2004introduction}. For lead separation $a$, the nucleation of a PSC will cause a voltage drop that exceeds its normal state value when $a < \Lambda_Q$. When the bias current or temperature is increased, another PSC can nucleate in the close vicinity, but the 'effective repulsion' between the PSCs, which now have to organize within a distance of $a/2$, will shrink the extent of the charge-imbalance region $\Lambda_Q$. Thus the excess voltage is reduced, which also creates sequential non-monotonic features, and negative differential resistance (NDR) spikes, in the $I-V$ characteristics \cite{strunk1998resistance,park1997resistance}.

%Another mechanism for excess resistance was introduced by () which suggests that the difference between the quasi-particle chemical potential ($\mu_{N}$) and the superconducting pair potential($\mu_{S}$) across S-N interfaces or phase slip centres (PSC) can lead to excess voltage. When a current bias is applied through a S-N interface or a PSC, it forms a charge-imbalance region due the non-equilibrium nature of the electrochemical potentials. The extension of charge-imbalance region is characterized by a length-scale $\Lambda_Q$ over which $\mu_{N}$ shows an exponential decay. In contrast, $\mu_{S}$ is constant within the S-regions but  shows a discontinuity at the PSC. Near $T_C$, $\Lambda_Q$ enhances significantly and can be $\Lambda_Q > \xi $, where $\xi$ is the coherence length. If the voltage-probe separation $a < \Lambda_Q$, then the pair of S-probes register a higher voltage than the N-probes as displayed by the schematics for S-N interface and a PSC in Fig.~2b. 

We note that the above model is valid under two crucial assumptions: (1) voltage-probe separation, $a < \Lambda_Q$ and (2) voltage probes are superconducting. The validity of these two criteria may seem unlikely in the Ag/Au nanostructure films. However, recent investigation of the $I-V$ characteristics in these systems have indicated possible formation of PSCs at the resistive transition \cite{islam2019current}, and attributed to a highly inhomogeneous and filamentary network of superconducting channels. Both resistance anisotropy and the strong spatial dependence of $R_{max}/R_N$ ratio (Fig.~1c inset) support this scenario. The microscopic superconducting filaments can form the voltage probes to the S-N interfaces in the current path in a fortuitous network pattern, and the condition of $a < \Lambda_Q$ can be satisfied locally (schematic in Fig.~3a).
%Owing to the non-equilibrium nature of charge-imbalance region, it is observed that the excess is independent of the bias current for lower currents and near critical current $I_C$, it decreases with increasing $I$. This led us to investigate the resistance peak with varying DC bias current. As shown in Fig.~2c, the $R/R_N$ is almost independent of $I$ within the range of $10$~$ \mu$A to $7$~mA. This may be attributed to the high value of $I_C$ near the transition as discussed in the earlier reports in Ag/Au nano-structures.

In order to explore the possibility of sequential excitation of PSCs in our case, we have investigated the 4-probe $I-V$ characteristics of the $R(12,78)$ channel close to the resistive transition during cooling in a separate thermal cycle. The $I-V$ measurement consists of three ramping segments while cooling at $\sim 10$~mK/min: (i) ramping of current to $-10$~mA, (ii) subsequently ramping up to $+10$~mA, and (iii) finally decreasing down to $0$~mA. During each $I-V$ cycle, $T$ decreases by $\approx 20$~mK. Fig.~3c shows the increasing segment ($|I|: 0 \rightarrow 10$~mA) of the $I-V$ characteristics at four temperatures across the transition. We first obtain the $T$-dependence of $R$ from the four-probe voltage at $-1$~mA in each $I-V$ cycle (Fig.~3b). From the thermally activated phase slip (TAPS) model \cite{islam2019current,chu2004phase} fit to the $R-T$ data, we obtain $T_C \approx 254.2$~K in this cooling cycle. Notably, we do not observe a clear ERA in this cycle, possibly due to the indirect process followed to extract the $R-T$ data from $I-V$ characteristics.

\begin{figure}[t]
\includegraphics[clip,width=9cm]{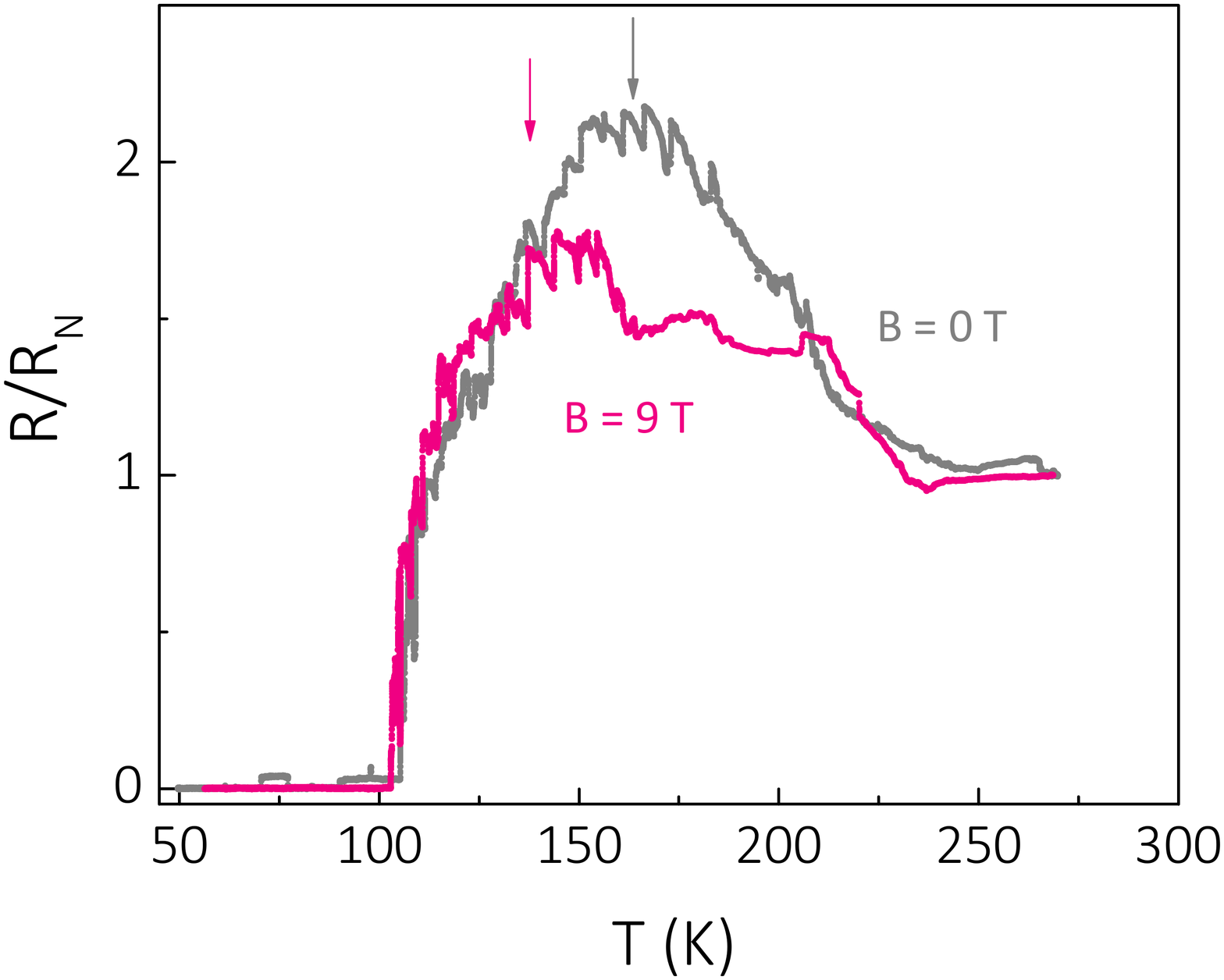}\caption{\textbf{Magnetic field dependence of resistance anomaly}: Comparison of $R-T$ plot at $0$~T and $9$~T perpendicular magnetic field in Ag/Au nanostructure film.  At high $T$, the anomaly is partially quenched at $9$~T, and the resistance maximum is shifted to lower $T$ (see text).}
\end{figure} 

The $I-V$ characteristics in Fig.~3c clearly exhibit a $T$-dependent critical current $I_C$, symmetric in both directions, above which the voltage drop increases rapidly. We note: (1) As $T$ is lowered, the $I-V$s becomes strongly hysteretic between the increasing and decreasing segments of $|I|$ (Inset of Fig.~3c, and supplementary information), as reported earlier in such films \cite{islam2019current}. Such hysteresis is often observed in quasi-1D superconductors when driven above $I_C$, and attributed to localized hot spots which involves Joule heating of dissipative centres like PSCs \cite{skocpol1974phase}. (2) For $I \gtrsim I_C$, $V$ increases in small abrupt jumps which can be attributed to the sequential introduction of PSCs thereby a discontinuous change in the chemical potential. (3) Determining $I_C$ by extrapolating the high-current regime (dashed lines in Fig.~3c), we find that $I_C \propto (T_C - T)$ (Fig.~3d), which deviates from the expected dependence of $I_C \propto (T_C -T)^{3/2}$ (dotted line in Fig.~3d) in the Ginzburg-Landau theory \cite{tinkham2004introduction}, but consistent with previous report on such films \cite{islam2019current}.    

Intriguingly, when $T$ is lowered below $\sim 70$~mK from $T_C$, the $I-V$s exhibit a new feature as shown in Fig.~3e. While the increasing segment of current up to $I \sim I_C$ ($\gtrsim 10$~mA) remains mostly featureless, the four-probe $V$ exhibits sharp peaks as $|I|$ is reduced during the decreasing segment ($|I|: 10 \rightarrow 0$~mA). The sharp peaks gradually disappear when $T$ is reduced $\sim 130$~mK below the $T_C$. Similar sequence of peaks in $I-V$ characteristics were observed during controlled heating as well (See supplementary information). The positions of the peaks are nearly symmetric in regard to the direction of current ($\sim \pm 4.5, 6.3$~mA etc.), although the peak voltage magnitude is not so, presumably due to the small temperature difference $\approx 20$~mK between the two ramping segments. The sequence of peaks in $I-V$ results in NDR as shown in Fig.~3f for three representative $T$. The non-monotonic structures and NDR at zero and finite current/voltage bias has been observed for wide variety of superconducting systems and can arise from different mechanisms such as Coulomb blockade of Cooper-pair tunneling in Josephson-coupled junctions \cite{watanabe2001coulomb} or interaction between the non-equilibrium regions around PSCs \cite{park1995resistance}. The coherent tunneling of Cooper pairs in superconducting tunnel-junction is however a low temperature quantum effect and not necessarily confined to only the fluctuation regime near $T_C$. Following the previous observation of excess voltage near $T_C$, the interacting non-equilibrium regions across S-N interfaces or PSCs may be responsible for the sequence of peaks in $I-V$ and subsequently NDR in our Ag/Au nanostructure film. We note that the same is applicable for charge-imbalance regions with multiple S-N interfaces. The observation of these peaks only in the decreasing current segment is not fully understood (Fig.~3e), although we anticipate the likelihood of thermally activated nucleation of PSCs according to Langer-Ambegaokar-McCumber-Halperin (LAMH) mechanism \cite{langer1967intrinsic,mccumber1970time}, once the local hotspots are created by driving the current to a high value ($\sim 10$~mA) during the increasing segment.

Finally, we discuss the effect of (perpendicular) magnetic field on the ERA using a Ag/Au nanostructure film, with lower $T_C \sim 100$~K but a rather broad ($\sim 100$~K) excess resistance region (Fig.~4), possibly due to greater level of inhomogeneity. We compare two cooling cycles of this film at fixed magnetic fields of zero and 9 Tesla. While the $T_C$ itself showed very little ($\approx 2.5$~K) decrease at $9$~T, which is consistent with earlier suggestion of a short coherence length \cite{islam2019current}, we observe (1) $R_{max}/R_N$ ratio is (partially) quenched in the presence of magnetic field at higher $T$ ($\geq 130$~K), and (2) the maximum of $R$ shifts to lower $T$ (shown by the vertical arrows in Fig.~4). Both observations are consistent with the previous reports on effect of magnetic field on the excess resistance anomaly \cite{arutyunov1999resistive,lindqvist1990new,kwong1991interfacial,crusellas1992giant}. Within the charge imbalance model, the suppression of excess resistance at finite field is linked to the decrease in $\Lambda_Q$ though orbital pair breaking and spin-flip scattering processes \cite{strunk1998resistance}, whereas argument regarding negative magneto-resistance for the anomaly is based on the enhanced proximity effects that drives the charge-imbalance region away from the voltage probes \cite{kwong1991interfacial}. However, other mechanisms such as interaction between the superconducting fluctuations and the conduction electrons \cite{lindqvist1990new,gordon1986electron} as well as the presence of collective modes in $1$D superconducting filaments \cite{tidecks2006current} can also reduce the resistance anomaly in the presence a magnetic field but these possibilities are probably unrelated to our observations in Ag/Au nano-structures.

In conclusion, we report the observation of resistance anomaly above the normal-state value in different regions of the Ag/Au nanostructure thin film near $T_C$. The anomalous resistance persists over a range of bias currents and the thermal history of the sample. The $I-V$ characteristics exhibits critical current behavior, and current-driven sequential voltage peaks for $T < T_C$, which manifests as negative differential resistances. Finally, we show that the resistance anomaly is partially quenched by the application of a perpendicular field of $9$~T. These observations serve as strong evidence of formation normal-superconductor interfaces in Ag/Au films at temperatures close to room temperature.

\bibliographystyle{naturemag}
\bibliography{RTS}

\subsection{Supplementary Information}
\begin{figure*}[t]
\includegraphics[clip,width=13cm]{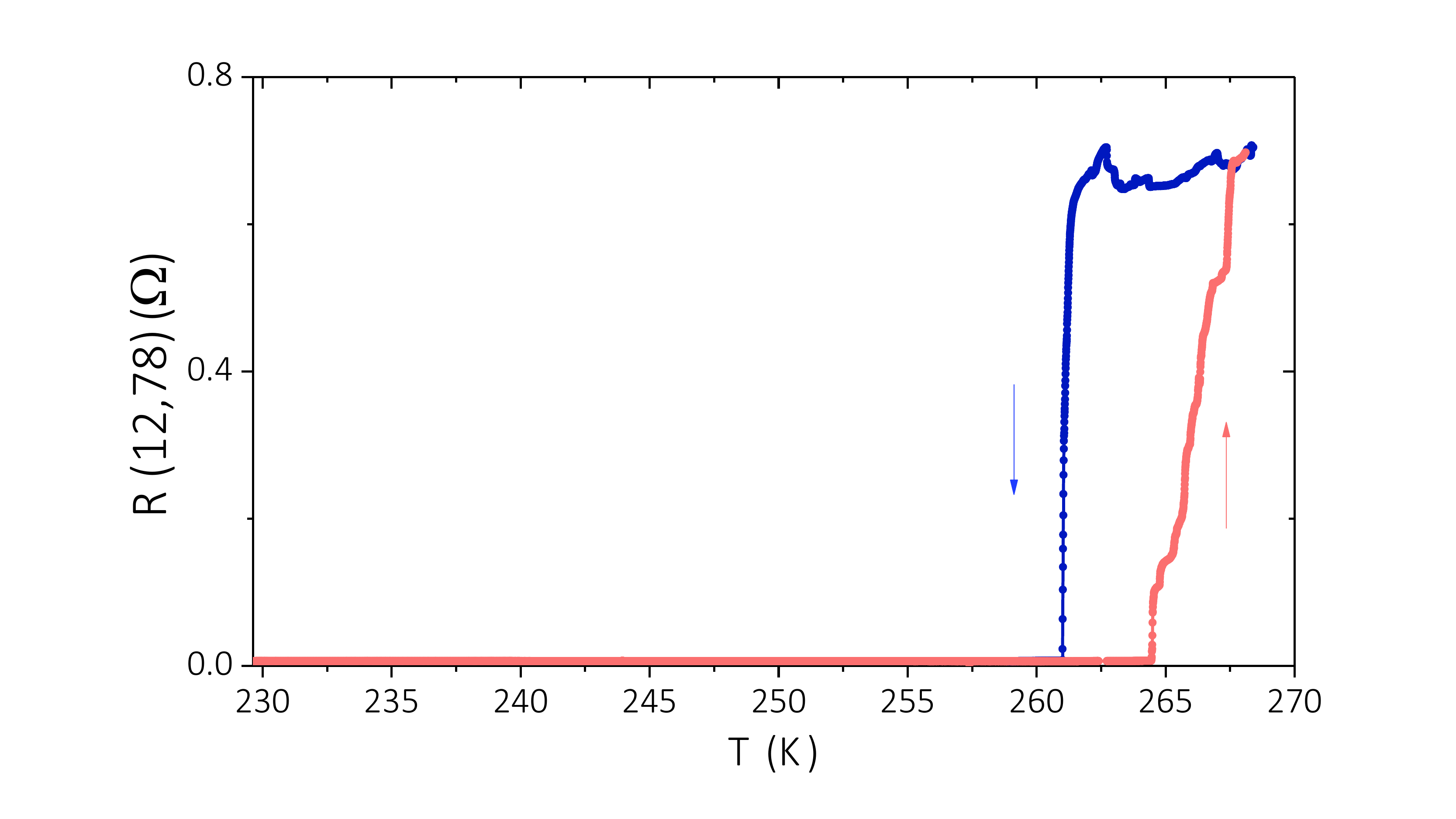}\caption{\textbf{Hysteresis in $T_C$ between cooling and heating cycle}: $R-T$ plots and the hysteresis of $T_C$ between the cooling cycle and the immediate heating cycle at a fixed ramping rate of $83$~mK/min.}
\end{figure*}

\begin{figure*}[t]
\includegraphics[clip,width=10cm]{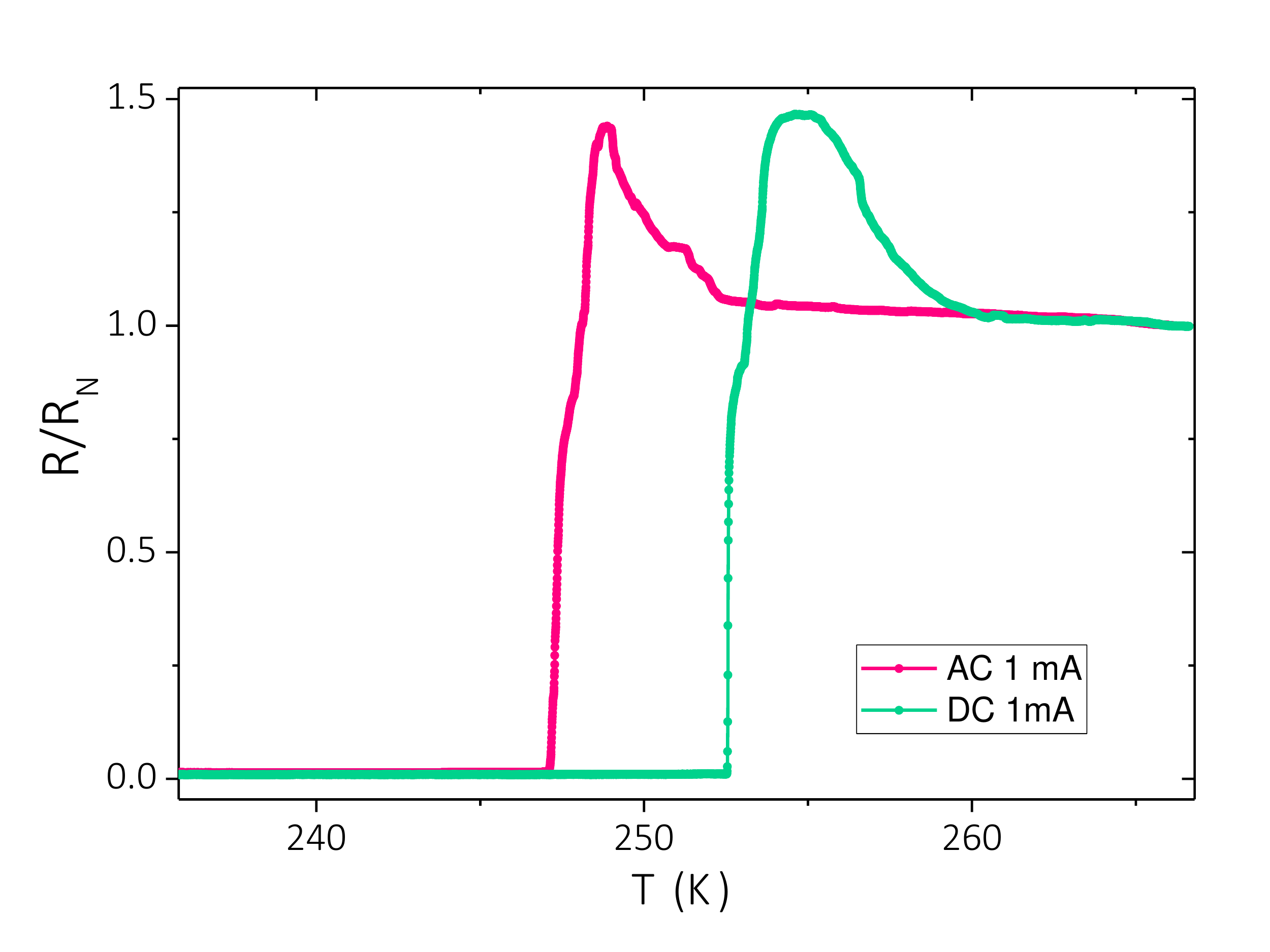}\caption{\textbf{ Excess resistance anomaly in AC and DC current biasing}: The anomalous increase in resistance was observed irrespective of AC and DC current biasing. The $R-T$ curves for both current biasing show similar maximum $R/R_N$ ratio.}
\end{figure*}

\begin{figure*}[t]
\includegraphics[clip,width=12.5cm]{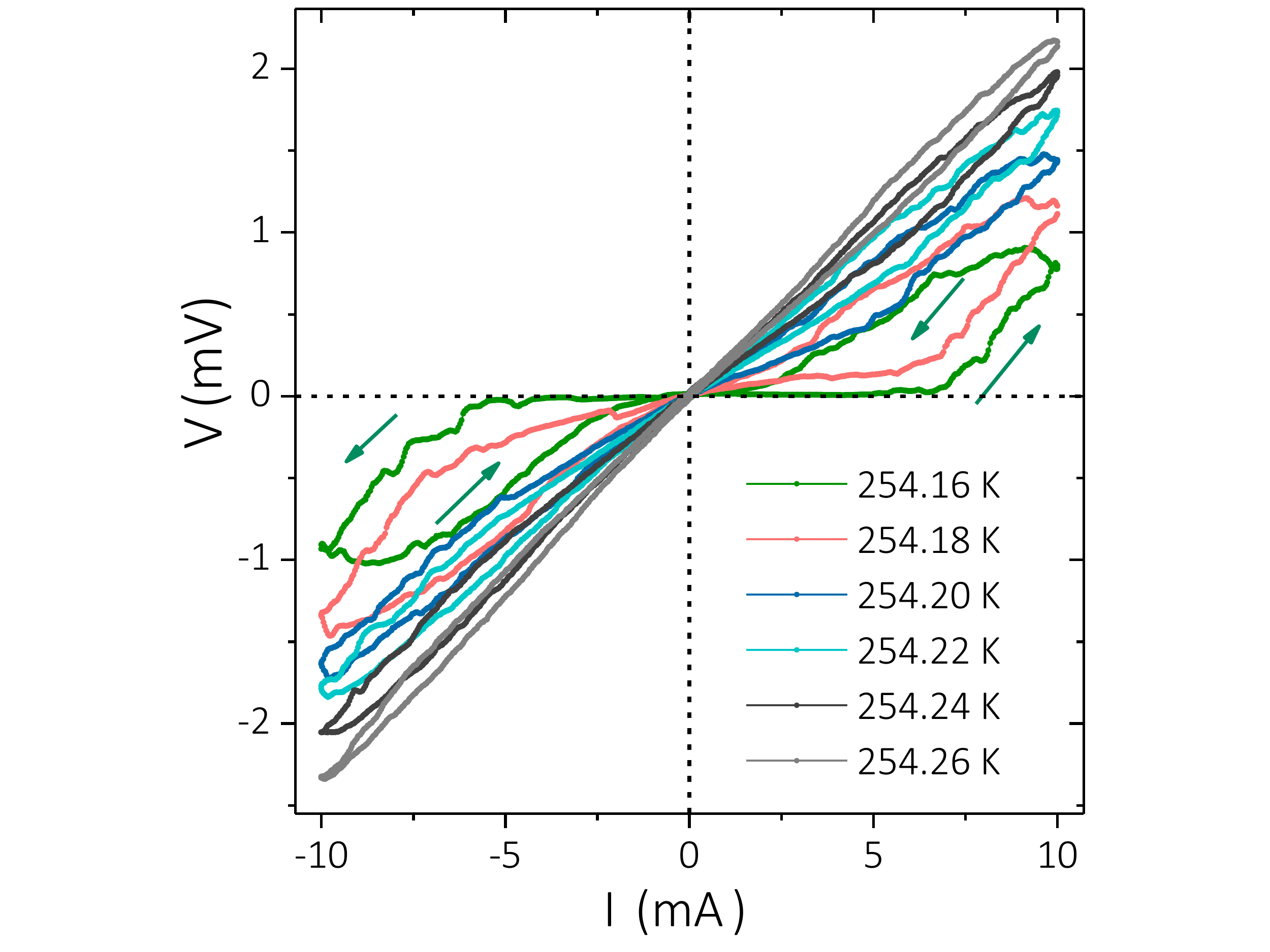}\caption{\textbf{Hysteresis in $I-V$ characteristics}: $I-V$ cycles taken across the transition for six different temperatures during a controlled cooling cycle. The ramping segments with increasing/decreasing $I$ show greater hysteresis at lower $T$. Similar hysteretic $I-V$ characteristics are observed in earlier reports of such films \cite{islam2019current}, and attributed to local heating effects in the dissipative points like phase slip centres.}
\end{figure*}

\begin{figure*}[t]
\includegraphics[clip,width=16cm]{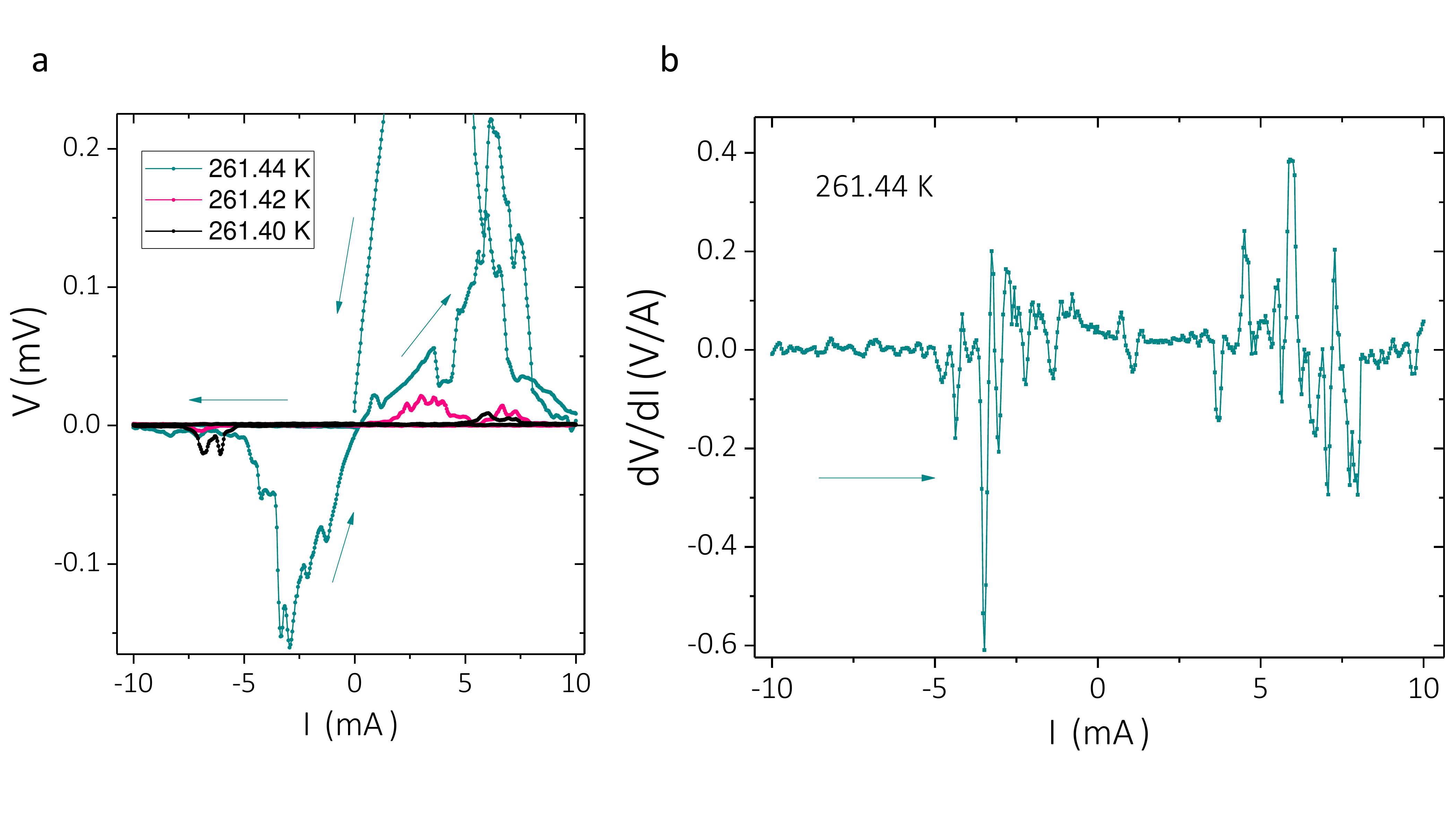}\caption{\textbf{Sequence of peaks in $I-V$ cycles and negative differential resistance (during heating) }: As mentioned in the main text, we observe similar sequence of sharp voltage peaks in $I-V$ loops taken during a heating cycle. The $I-V$ curves for three different temperatures are shown in (a). (b) The negative differential resistance obtained for $T = 261.44$~K shown for the current ramping segment $-10$~mA to $+10$~mA.}
\end{figure*}

\end{document}